\documentclass[journal, draftcls, one column, 12pt]{IEEEtran}
\usepackage{amsmath}
\usepackage{graphicx}
\usepackage{amssymb}
\usepackage{cases}
\usepackage{cite}
\usepackage[ruled,vlined,lined,ruled,linesnumbered]{algorithm2e}
\usepackage{relsize}
\usepackage{multirow}
\usepackage[nodisplayskipstretch]{setspace}
\usepackage[top=1in, bottom=1in, left=0.7in, right=0.7in]{geometry}
\usepackage[usenames, dvipsnames]{color}
\usepackage{epstopdf}

\begin{document}
\title{Wireless Communications in the Era of Big Data}
\author{Suzhi~Bi, Rui Zhang, Zhi Ding, and Shuguang Cui
\thanks{S.~Bi and R.~Zhang are with the  Department of Electrical and Computer Engineering, National University of Singapore, Singapore (E-mail:\{bsz,~elezhang\}@nus.edu.sg).}
        \thanks{Z.~Ding is with the  Department of Electrical and Computer Engineering, University of California, Davis, CA, USA (Email: zding@ucdavis.edu).}
    \thanks{S.~Cui is with the Department of Electrical and Computer Engineering, Texas A$\&$M University, College Station, TX, USA (Email: cui@ece.tamu.edu). He is also a Distinguished Adjunct Professor at King Abdulaziz University in Saudi Arabia.}}
\maketitle

\vspace{-1.8cm}

\section*{Abstract}
The rapidly growing wave of wireless data service is pushing against the boundary of our communication network's processing power. The pervasive and exponentially increasing data traffic present imminent challenges to all the aspects of the wireless system design, such as spectrum efficiency, computing capabilities and fronthaul/backhaul link capacity. In this article, we discuss the challenges and opportunities in the design of scalable wireless systems to embrace such a ``bigdata" era. On one hand, we review the state-of-the-art networking architectures and signal processing techniques adaptable for managing the bigdata traffic in wireless networks. On the other hand, instead of viewing mobile bigdata as a unwanted burden, we introduce methods to capitalize from the vast data traffic, for building a bigdata-aware wireless network with better wireless service quality and new mobile applications. We highlight several promising future research directions for wireless communications in the mobile bigdata era.

\section{Introduction}
\begin{figure}
\centering
  \begin{center}
    \includegraphics[width=0.7\textwidth]{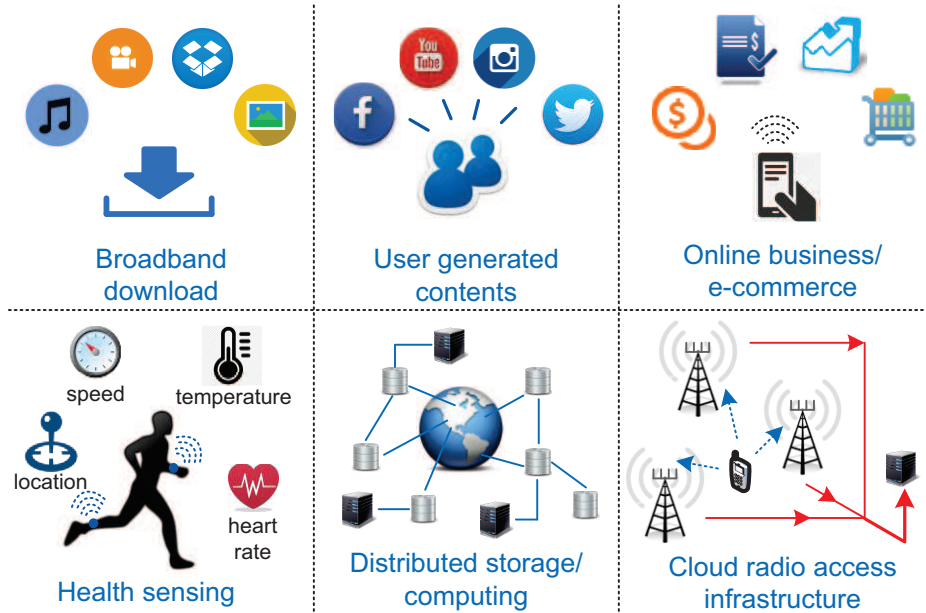}
  \end{center}
  \caption{Some example sources of wireless bigdata traffic.}
  \label{71}
\end{figure}

Decades of exponential growth in commercial data services has ushered in the so-called ``bigdata" era, to which the expansive mobile wireless network is a critical data contributor. As of $2014$, the global penetration of mobile subscribers has reached $97\%$, producing staggeringly $10.7$ ExaBytes ($10.7 \times 10^{18}$) of mobile data worldwide. The surge of mobile data traffic in recent years is mainly attributed to the popularity of smartphones, phone cameras, mobile tablets and other smart mobile devices that support mobile broadband applications, e.g., online music, video and gaming as shown in Fig.~\ref{71}. With a compound annual growth rate of over $40\%$, it is expected that the mobile data traffic will increase by $5$ times from $2015$ to $2020$.

In addition to the vast amount of wireless source data, modern wireless signal processing often amplifies the system's pressure from bigdata in pursuit of higher performance gain. For instance, MIMO antenna technologies are now extensively used to boost throughput and reliability at both mobile terminals (MTs) and base stations (BSs) of high speed wireless services. This, however, also increases the system data traffic to be processed in proportion to the number of antennas in use. Moreover, the $5$G (the fifth generation) wireless network presently under development is likely to migrate the currently hierarchical, BS-centric cellular architecture to a cloud-based layered network structure, consisting of a large number of cooperating wireless access points (APs) connected by either wireline or wireless fronthaul links to a bigdata capable processing central unit (CU). New wireless access structures, such as coordinated multipoint (CoMP or networked MIMO) \cite{2011:Irmer}, heterogeneous network (HetNet) \cite{2013:Andrews} and cloud-based radio access network (C-RAN) \cite{2013:Park1}, are under development to achieve multi-standard, interference-aware and energy-friendly (green) wireless communications. In practice, the use of cooperating wireless APs could easily generate multiple Gbps data from a single user's fronthaul links due to the need for baseband joint processing, such that the high traffic load may overwhelm the fronthaul link or the system computing unit for signal processing and coordination. Such intensely high system traffic volume, together with the rapidly growing mobile data source volume, surpasses both the processing power improvement speed of our current computing capabilities and the fronthaul/backhaul link rate increase pace of our networking systems. It necessitates a new wireless architecture along with efficient signal processing methods to make wireless systems \emph{scalable} to continued growth of data traffic.

On the other hand, timely and cost-efficient information processing is made possible by the fact that the vast-volume mobile data traffics are not completely chaotic and hopelessly beyond management. Rather, they often exhibit strong \textit{insightful features}, such as user mobility pattern, spatial, temporal and social correlations of data contents. These special characteristics of mobile traffic present us with opportunities to harness and exploit bigdata for potential performance gains in various wireless services. To effectively utilize and exploit these characteristics, they should be identified, extracted, and efficiently stored. For instance, caching popular contents at wireless hot spots could effectively reduce the real-time traffic in the fronthaul links. Additionally, network control decisions, such as routing, resource allocation, and status reporting, instead of being rigidly programmed, could be made data-driven to fully capture the interplay between bigdata and network structure. Presently, however, these advanced data-aware features could not be efficiently implemented in current wireless systems, which are mainly designed for content delivery, instead of analyzing and making use of the data traffic.

Bearing in mind of the aforementioned challenges and opportunities brought by bigdata traffic, we address in this article two important problems of wireless communication system design in the bigdata era:
\begin{itemize}
  \item[\textbf{Q1}:] What may constitute a \emph{scalable} wireless network architecture for efficient handling of bigdata traffic?
  \item[\textbf{Q2}:] How to effectively incorporate and utilize the \emph{bigdata awareness} to improve the wireless system performance?
\end{itemize}
Specifically, to answer \textbf{Q1}, we introduce in Section \ref{sec:scalable_structure} a hybrid signal processing paradigm to enable flexible data processing at both the BS/AP and the CU levels, and correspondingly a number of scalable data traffic management techniques to serve the conflicting needs between the overall system performance and the data processing complexity. For \textbf{Q2}, we first discuss in Section \ref{sec:capital_big_data} typical bigdata features and efficient data analytics to extract these features. Next, we introduce a number of bigdata-aware signal processing methods and wireless networking structures to capitalize from bigdata interplay, such as mobile cloud processing, crowd computing, and software-defined networking, etc. We also suggest in Section \ref{sec:future} several future research directions for wireless communications in the bigdata era. Finally, we conclude this article in Section \ref{sec:conclusion}.

Before proceeding to detailed discussions, it is worth mentioning that the considered scalable network structure and bigdata awareness are both important mechanisms for accommodating mobile bigdata in future wireless networks, though they focus differently on the physical and network/application layers, respectively. Nonetheless, the two solutions can also be complementary to each other. For instance, as we will discuss later, we can optimize the overall caching strategy by combining long-term cache provisioning (network/application layer) and real-time cache-assisted signal processing (physical layer) techniques. In addition, although this article focuses on the design aspects of cellular networks in the bigdata era, most of the key enabling mechanisms for mobile bigdata processing are also applicable to other wireless networking structures, such as wireless local area networks (WLANs) and heterogeneous networks. Some representative system designs are also discussed in this article.

\section{Scalable Wireless Bigdata Traffic Management}\label{sec:scalable_structure}
\subsection{A hybrid network structure}
Neither the current cellular systems nor the next-generation cloud-based C-RAN \cite{2013:Park1} under development was designed to provide a scalable solution for the arrival of the bigdata era. The current $3$G and $4$G cellular systems exemplify a BS-centric design, in which a BS bears much the responsibilities of radio access, baseband processing and radio resource control execution to serve the mobile users in the vicinity. To meet the fast growing mobile data service demand, a smaller cell size is commonly used to improve frequency reuse, which may generate complex and severe inter-cell interference. Furthermore, small cells can also be costly because of cost from densely deployed BSs. The cloud-centric network proposed for $5$G mitigates the inter-cell interference by centralized signal processing and reduces the unit cell deployment cost by moving computations to the ``cloud". At the same time, only inexpensive relay-like remote radio heads (RRHs) are used for radio frequency (RF) level wireless access. However, such fully centralized scheme may be overwhelmed by the huge wave of data traffic beyond its fronthaul link capacities and its computational power.

\begin{figure}
\centering
  \begin{center}
    \includegraphics[width=0.7\textwidth]{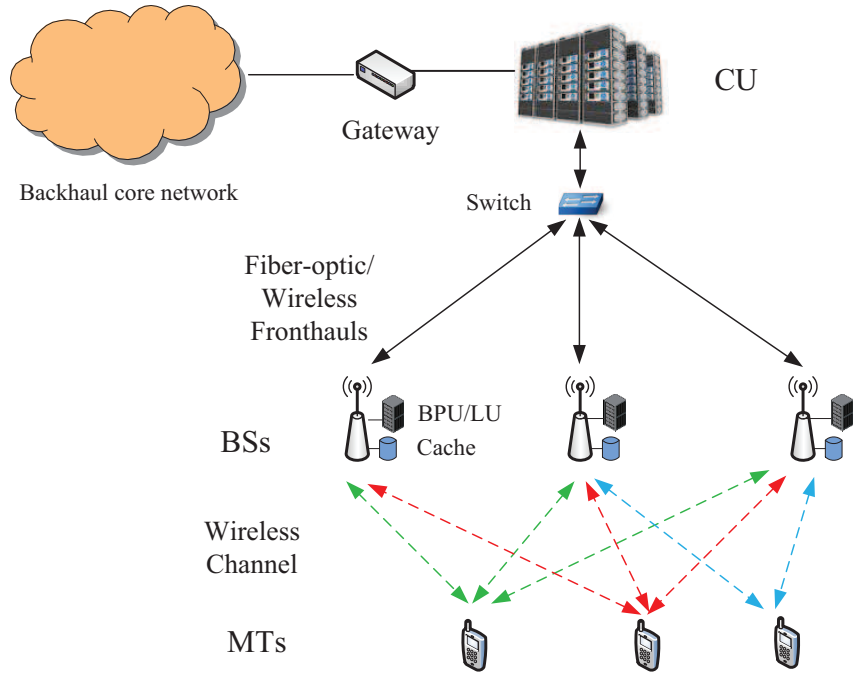}
  \end{center}
  \caption{A hybrid CU-BS processing network structure.}
  \label{72}
\end{figure}

Alternatively, a \emph{hybrid} structure could take advantage of the benefits from the two design paradigms; that is, a wireless system that could adaptively choose only local processing at the BS-level, or only central processing at the CU-level, or parallel processing at both levels, based on, for instance, physical channel conditions and correlations in the data contents, etc. We thus consider such a generic network structure shown in Fig.~\ref{72}, which mainly inherits the skeleton of C-RAN, but has integrated several programmable modules to carry out intelligent signal processing at the BS level. In the radio access network, mobile users could be served simultaneously by multiple BSs, where each BS is equipped with multiple antennas and is linked to the CU via high-speed fiber/wireless fronthauls for exchanging user data and control signals. The CU is further connected to the backhaul core network for external content access. In the proposed hybrid network structure, baseband processing units (BPUs) are available at both the BSs and CU, which enable user message encoding/decoding at both levels. In addition, learning units (LUs) are installed for data traffic analytics, whose functions will be detailed in Section \ref{sec:capital_big_data}. Caches are also installed at the BSs and CU to save the fronthaul bandwidth consumed for frequent retransmissions of popular contents.

Before entering the discussions of hybrid signal processing models, it is worth mentioning that applicable fronthaul data management methods are directly constrained by fronthaul technologies in use. Specifically, the system could choose between optical analog and optical/wireless digital fronthaul technologies. Optical analog modulation using radio frequency (RF) signal as input is commonly referred to as the radio-over-fiber (RoF). Alternatively, analog RF input signal could also be quantized and encoded into binary codewords for digital wireless or fiber-optic communication (DFC). In practice, RoF is simpler and less expensive than DFC. Furthermore, it also exhibits lower processing delays and better interoperability with multiple wireless standards, e.g., $3$G, LTE and WiFi, as it is oblivious to the user codebooks and wireless modulation schemes. However, its limitations are also evident, e.g., susceptibility to noise and signal distortion, and difficulty of synchronization. More importantly, available signal processing techniques for fronthaul traffic management using RoF are less sophisticated, generally limited to simply transforming, or removing certain parts of the received RF signals, e.g., sub-channel and antenna selection methods. In contrast, DFC could be combined with data compression, opportunistic decoding and many other advanced digital signal processing techniques. In the following, we mainly focus on data traffic management methods using digital fronthaul.

\subsection{Hybrid signal processing models}\label{sec:hybrid}
The wireless/fiber-optic link has its own throughput limit. For instance, a commercial fiber-optic link normally operates at a link rate in the order of $10$ Gbps for digital communication over a single optical carrier. Transmission rates beyond the link rate capacity may lead to  severe signal distortions, and consequently poor decoding performance. Therefore, the system performance must be optimized under the fronthaul link capacity constraints. With respect to the hybrid network structure in Fig.~$\ref{72}$, we now introduce some scalable fronthaul data management techniques in three major categories:
\subsubsection{Data compression}
Uplink direction would require unlimited fronthaul capacity to transmit an analog RF signal perfectly without any distortion from a BS to the CU. An analog signal could be more efficiently transmitted through the fronthaul if it is quantized and compressed into binary codewords. From an information theoretic perspective, the effect of data compression could be modeled as a test channel (often Gaussian for simplicity of analysis) for which uncompressed signals as the input and compressed signals are the output. The compression design is equivalent to setting the variance of the additive compression noise \cite{2009:Sanderovich}. To achieve successful compression, the encoder needs to transmit to the decoder at a rate at least equal to the mutual information between the input and the output over the Gaussian test channel. Intuitively, a tighter fronthaul capacity constraint would therefore require a more ``coarse compression" with a larger compression noise. Existing compression designs in general take the following approaches.

\begin{itemize}
  \item \emph{Joint compression across different BSs: } When multiple BSs compress and forward their received signals to the CU in uplink, the compression design requires setting the covariance of the compression noises across different BSs. A common objective is to maximize the information rate under the fronthaul capacity constraints. In this setting, \textit{distributed Wyner-Ziv lossy compression} can be used at the BSs, exploiting signal correlation across the multiple BSs \cite{2009:Sanderovich}. The distributed Wyner-Ziv compression scheme is shown to yield significant capacity gains over independent quantization methods especially in the low backhaul capacity region \cite{2009:Sanderovich}. Similar data compression methods could also be applied in downlink. Interestingly, it has been shown in \cite{2013:Park1} that downlink compression and multi-user precoding design (for interference mitigation) could be designed separately without compromising maximum system throughput, which is achieved by an optimal but much more complicated joint compression-precoding design.
  \item \emph{Independent BS-level compression:} The practical implementation of distributed Wyner-Ziv compression is difficult mainly because of the high complexity in determining the optimal joint compression codebook and the joint decompressing/decoding at the CU. Accordingly, independent compression methods, where the quantization codebook at a BS is only determined by its local signal-to-noise ratio (SNR), can be used to reduce the computational complexity and the signaling exchange overhead in the fronthaul.
  \item \emph{Uniform scalar quantization:} Even when using independent BS-level compression, real-time computation and exchange of quantization codebooks using the information-theoretical source coding approaches are often difficult to realize in practice. Instead, simple uniform scalar quantization methods compatible with A/D modules are proposed to reduce the implementation cost \cite{2014:Liu1}. Interestingly, it is shown in \cite{2014:Liu1} that the achievable rate using simple uniform scalar quantization in fact performs closely to that of the Gaussian test channel model. This indicates that efficient fronthaul capacity usage is achievable in practical systems with simple quantization methods.
\end{itemize}

\subsubsection{BS-level encoding/decoding}
Besides acting as relays to compress/decompress and forward the user signals, BSs with advanced baseband processing capabilities could also encode/decode the received messages to further improve the system performance under stringent fronthaul capacity constraints.
\begin{itemize}
  \item \emph{Partial cooperation: } In uplink, one direct method to reduce fronthaul traffic is to limit the number of cooperating elements when serving mobile users. Many sparsity inducing optimization methods could be applied to satisfy a certain quality of service level using minimum numbers of sub-channels, antennas or cooperating BSs. In downlink, similar \textit{sparse precoding} methods could be studied to optimize precoders by jointly maximizing the user utilities (e.g., data rate) and minimizing the total number of data streams in the fronthaul \cite{2013:Hong}.
  \item \emph{Distributed encoding/decoding: } Distributed decoding allows the BSs to decode user messages locally without forwarding quantized signals to the CU. For instance, \cite{2009:Sanderovich} considers a rate-splitting approach to divide an MT's message into two parts, where one part is decoded locally by the serving BS and the remainder is compressed and jointly decoded by the CU. In another case, \cite{2011:Zhang} proposes an opportunistic hybrid decoding method, where a user's message is either decoded locally at a BS when its SNR is sufficiently high, or jointly decoded by the CU based on signals forwarded from a subset of cooperating BSs when the SNR at each individual BS is too low. Note that the locally decoded user messages can be used to cancel their interferences to the received RF signals at the BSs, which can effectively reduce the amount of data transmitted to the CU over the fronthaul links. In the downlink case, BSs could \textit{encode and modulate} the baseband symbols to RF signals before transmitting them to the MTs. Therefore, instead of transmitting complete signal waveforms (or waveform samples) to the BSs, CU could save fronthaul bandwidth by transmitting separately the information symbols and the beamforming vectors, while leaving RF modulation to the BSs.
\end{itemize}

\begin{figure}
\centering
  \begin{center}
    \includegraphics[width=0.9\textwidth]{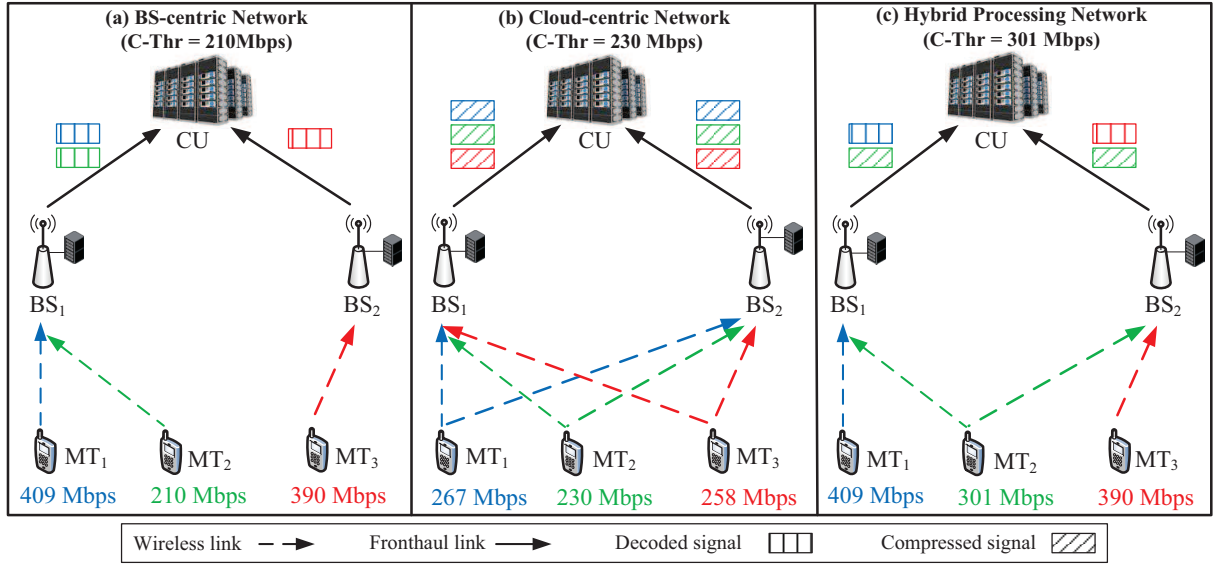}
  \end{center}
  \caption{Throughput performance comparison of three network structures (from left to right): BS-centric, cloud-centric, and hybrid processing  networks. The common-throughput (C-Thr) achieved by the three networks from (a) to (c) are $210$, $230$, and $301$ Mbps, respectively.}
  \label{73}
\end{figure}

To show the performance advantage of the hybrid signal processing model, we present a numerical example in Fig.~\ref{73} to compare the throughput performance among the BS-centric, the cloud-centric, and the hybrid processing networks. Let us consider a cellular uplink, where $3$ MTs transmit over orthogonal sub-channels, each with $100$ MHz bandwidth. Besides, each fronthaul link has $1.2$ Gbps capacity. The decoding methods of the three networks are described as follows.
\begin{itemize}
  \item BS-centric network: BS$_1$ decodes the messages from MT$_1$ and MT$_2$, and BS$_2$ decodes the message from MT$_3$. Then both the BSs send the decoded messages to the CU;
  \item Cloud-centric network: both BSs compress the received signals using the scalar quantization method considered in \cite{2014:Liu1}. They then forward the compressed signals to the CU for joint decoding. In particular, each user is equally allocated $400$ Mbps fronthaul bandwidth at a BS to transmit its compressed signal;
  \item Hybrid processing network: BS$_1$ and BS$_2$ first decode the messages from MT$_1$ and MT$_3$, respectively, before transmitting the decoded messages to the CU. Meanwhile, each BS uses the remaining fronthaul bandwidth to compress and transmit the signal from MT$_2$ to the CU for the joint decoding of MT$_2$'s message.
\end{itemize}
From the aforementioned network setups, we calculate in Fig.~\ref{73} the achievable user data rates under a random channel realization, and compare the common-throughput performance (the minimum data rate among the three users) in different cases. We can see that the BS-centric network achieves the lowest common-throughput, owing to the low data rate of the cell-edge user MT$_2$, which is only $210$ Mbps. The cloud-centric network slightly improves the data rate of MT$_2$ and hence the common-throughput to $230$ Mbps, thanks to its joint processing gain. However, the data rates of MT$_1$ and MT$_3$ are severely degraded, since the limited fronthaul capacity introduces high compression noises to the useful signals. The hybrid processing network achieves the highest common-throughput ($301$ Mbps) among the three schemes that we considered, which is $43\%$ and $31\%$ higher than those of the BS-centric and cloud-centric networks, respectively. Compared to the cloud-centric networks, by decoding the messages from MT$_1$ and MT$_3$ at the BS-level, the hybrid processing network has a larger fronthaul bandwidth to spare for transmitting MT$_2$'s signals to the CU with more refined compression, thus achieving a higher joint processing gain.

\subsubsection{Cache-assisted processing}
\begin{figure}
\centering
  \begin{center}
    \includegraphics[width=0.5\textwidth]{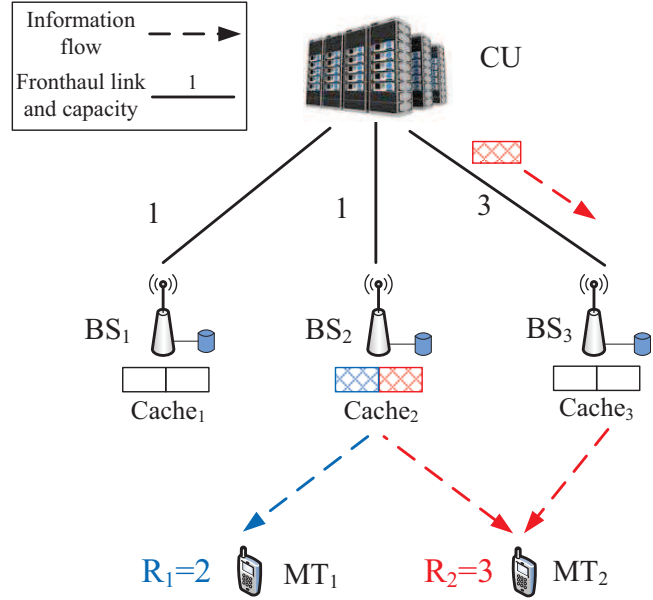}
  \end{center}
  \caption{Downlink cache-assisted wireless signal processing.}
  \label{74}
\end{figure}

In downlink transmission, caching at the BSs is cost-effective to reduce real-time traffic on fronthaul, thereby enabling significant improvement on the overall C-RAN performance. Cache-assisted wireless resource allocation is a cross-layer approach that incorporates the status of application-layer data flow in wireless physical-layer design. As an illustrative example in Fig.~\ref{74}, BS$_2$ serves two requests from the two MTs, whereas caches of the other two BSs are empty. Although MT$_1$ is closer to BS$_1$ with a better wireless channel condition, the maximum downlink data rate is only $1$ unit per second if BS$_1$ is selected to transmit directly, due to the constraint of link congestion between BS$_1$ and the CU. Instead, the CU could select BS$_2$ to send the cached contents to MT$_1$ at a rate of $2$ units per second, whose end-to-end data rate is not constrained by the congestion level of the CU-to-BS$_2$ link. On the other hand, MT$_2$ could be served by two cooperating BSs (BS$_2$ and BS$_3$) with an improved wireless channel gain from coordinated beamforming. In particular, the CU only needs to transmit the content requested by MT$_2$ to BS$_3$ before the cooperative transmissions of the two BSs. Thanks to such wireless cooperation, MT$_2$ could achieve a higher data rate at $3$ units per second.

In a more general setting, caches could be located at not only the BSs, but also the routers and the CU. Furthermore, distributed caching could also be adopted at MTs to allow mobile users to serve popular contents requested by nearby peer users in a device-to-device (D2D) manner. We could foresee that cache-assisted resource allocation method becomes a key enabling factor of significant bandwidth saving, since frequent overlapping of requested objects will occur as the volume of mobile traffic increases. However, it also becomes a more challenging problem to optimize system-wide resource allocation due to the interleaving among cache placement, wireless interference, routing, and the combinatorial nature of node selections in the wireless network. A more comprehensive understanding on the design tradeoff remains open for future study.

Another interesting topic on cache-assisted resource allocation is on \textit{cache provisioning} for popular contents to reduce the real-time backhaul traffic. In particular, cache provisioning addresses the questions of what, where and when to cache in the wireless infrastructure. In this case, accurate knowledge of the mobile user demand profiles is a key to efficient cache provisioning. The extraction of user demand profiles from mobile data traffic is performed by wireless bigdata analytics, which will be discussed in the next section.

\section{Developing a Bigdata Aware Wireless Network}\label{sec:capital_big_data}
Instead of viewing mobile bigdata as a pure burden, we investigate in this section the potential performance gain from developing a bigdata-aware intelligent wireless network. However, its efficient operation relies on the in-depth knowledge of the wireless bigdata traffic characteristics. As most of such characteristics are implicit, we first introduce data-analytical methods necessary to extract these bigdata features. We then discuss how to leverage these bigdata characteristics in designing wireless networks to capitalize from the mobile bigdata traffic.

\subsection{Useful mobile bigdata features and applications}
There is clearly a strong connection between wireless service usage and human behavioral patterns in the physical world. For this reason, wireless data traffic contains strong correlative and statistical features in various dimensions, such as time, location and the underlying social relationship, etc. On one hand, mobile traffic has strong \emph{aggregate features}. For instance, there exist severe load imbalances spatially and temporally, such that, presently, $10\%$ of ``popular" BSs carry about $50\%\sim 60\%$ traffic load. The peak traffic volume at a given location is much higher than the regular average. These aggregate features could be exploited to reduce real-time fronthaul/backhaul traffic and to improve wireless network efficiency. Example applications include: cell planning according to geographical data usage distribution, peak load shifting via load-dependent pricing, and cache provisioning based on aggregate demand profile, among others.

On the other hand, each mobile user's data usage profile also exhibits a unique set of \emph{individual features}, such as mobility pattern, preference of various data applications, and service quality requirements. For instance, a mobile user's trajectories often consist of a very limited number of frequent positions and quasi-repetitive patterns. Besides, the recent popularity of mobile social networking interconnects seemingly uncorrelated individual data usages into a unified social profile, thereby presenting a novel perspective to analyze the mobile traffic pattern. These individual and social features are useful for system operators to personalize and improve wireless service quality. Many intelligent data-aware services could be provided according to user profiles. Examples include resource reservation in handoff using location prediction, context-aware personal wireless service adaptation, and mobility-based routing and paging control.

\subsection{Bigdata analytical tools}
The ability to acquire, analyze, and exploit mobile traffic characteristics can be accomplished by specially designed learning units (LUs) installed at both the BSs and CUs (see Fig.~\ref{72}). Their core enabling factors are the embedded data-analytical algorithms. Some commonly used algorithms for wireless traffic analysis and their main applications to wireless communications are classified as follows and summarized in Table I.
\subsubsection{Stochastic modeling}
Stochastic modeling methods use probabilistic models to capture the \emph{explicit} features and dynamics of the data traffic. Commonly used stochastic models include: order-$K$ Markov model, hidden Markov model, geometric model, time series, linear/nonlinear random dynamic systems, etc. For example, Markov models and Kalman filters are widely used to predict user mobility and service requirements \cite{2004:Akyildiz}. The collected user data are often used for parameter estimation of stochastic models, such as estimating the transition probability matrix of a Markov chain.
\subsubsection{Data mining}
Data mining focuses on exploiting the \emph{implicit} structures in the mobile data sets. Also taking the mobility prediction problem as an example, individual user's mobility pattern could be extracted and discovered by finding the most frequent trajectory segments in the mobility log. Prediction could be made accordingly by matching the current trajectory to the mobility profile. Clustering is another useful technique to identify the different patterns in the data sets. It is widely used in context-aware mobile computing, where a mobile user's context and behavioral information, such as sleeping and working, are identified from wireless sensing data for providing context-related services \cite{2006:Krause}.
\subsubsection{Machine learning}
The main objective of machine learning is to establish functional relationship between input data and output actions, thus achieving \emph{auto-processing} capability for unseen patterns of data inputs. Among the many useful techniques in machine learning applied to wireless communications, classification (determining the type of input data) and regression analysis (data fitting) are two common methods, whose applications include context identification of mobile usage and prediction of traffic levels (classification), or fitting the distributions of trajectory length, mobile user location, and channel holding times (regression). Besides, reinforcement learning, such as Q-learning \cite{2012:Yau}, is useful for taking proper real-time actions to maximize certain long-term rewards. A typical example is making the handoff and admission control decision (action), given the current traffic load (state) and incoming new requests (event), in which the reward could be evaluated against the reduction of dropped calls or failed connections.

\begin{table}
\caption{Summary of common wireless bigdata analytic tools and example applications}
\centering
\small
\begin{tabular}{|c|c|c|}
\hline
\emph{Subjects} & \emph{Models/algorithms} & \emph{Example wireless applications}\\ \hline
\multirow{2}{*}{\textbf{Statistical modeling}} & Markov models, time series,  &   mobility prediction, resource provision,                \\
                                      & geometric models, Kalman filters     &   device association/handoff prediction                   \\ \hline
\multirow{2}{*}{\textbf{Data mining}} & \multirow{2}{*}{pattern matching, text compression,}  &   mobility prediction, social group clustering,               \\
                                      &       &   context-aware processing,  cache               \\
                                      & clustering, dimension reduction                                   &  management,  user profile management \\ \hline
\multirow{11}{*}{\textbf{Machine learning}}     &  classification algorithms,               & context identification, traffic prediction,\\
                                               &   neural network,            & fitting trajectory length, user location \\
                                               &   regression analysis,       &  and the channel holding time   \\ \cline{2-3}
                                               &  dimension reduction algorithms:            & user data compression/storage, traffic\\
                                               &  PCA, PARAFAC, Tucker3             & feature extraction, blind multiuser detection \\ \cline{2-3}
                                               &  Q-learning            & handoff and admission controls \\ \cline{2-3}
                                               &  \multirow{2}{*}{primal/dual decomposition, ADMM}              &  distributed routing/rate control \\
                                               &            &  and wireless resource allocation\\ \cline{2-3}
                                               &  online convex optimization,               &  on-line mobility predictions, handoffs,\\
                                               &  stochastic learning                       &  and resource provisioning                \\ \cline{2-3}
                                               &  active learning, deep learning            &  incomplete/complex mobile data processing  \\ \hline

\end{tabular}
\end{table}

\subsubsection{Large-scale data analytics}
Wireless bigdata poses many challenges to the aforementioned conventional data-analytical methods due to its high volume, large dimensionality, uneven data qualities, and the complex features therein. To improve signal processing efficiency, one can combine the following complexity reduction techniques with the conventional data analytical tools for large-scale data processing.
\begin{itemize}
  \item \textit{Distributed optimization algorithms}, such as primal/dual decomposition and alternating direction method of multipliers (ADMM), are very useful to decouple large-scale statistical learning problems into small subproblems for parallel computations so as to relieve both the computational burden at the CU and the bandwidth pressures to the fronthaul/backhaul links.
  \item \textit{Dimension reduction} methods are useful to reduce the data volume to be processed while capturing the key features of bigdata. Among various methods, principle component analysis (PCA), along its many variants, is the mostly used method today. In addition, tensor decomposition methods are also popular in mobile data processing, which seek to approximately represent a high-order multi-way array (tensor) as a linear combination of outer products of low-order tensors. By doing so, the hardware requirement and cost for storing the high-order arrays of mobile data could be reduced.
  \item \textit{Other advanced learning methods} could be used to handle incomplete or complex data sets. Interesting examples include active learning, which deals with partially labeled data set; online learning for responding in real-time to sequentially received data; stochastic learning that makes a decision periodically in each time interval; and deep learning for modeling complex behaviors contained in a data set.
\end{itemize}

\subsection{Bigdata aware wireless network}
\begin{figure}
\centering
  \begin{center}
    \includegraphics[width=0.8\textwidth]{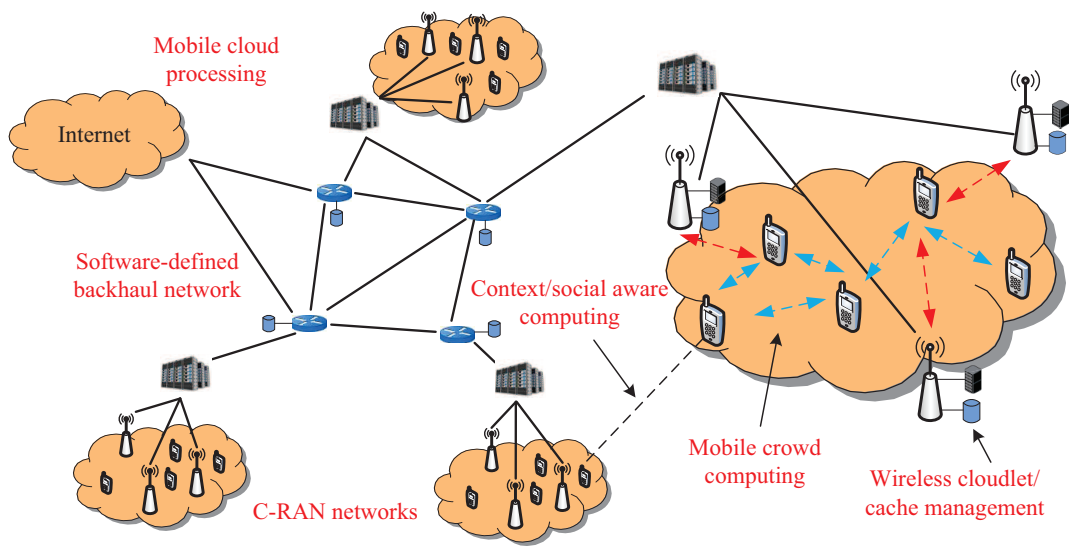}
  \end{center}
  \caption{An illustrative structure of bigdata aware wireless network.}
  \label{75}
\end{figure}

Once identified and extracted, data characteristics could be used to improve wireless service quality and generate new mobile applications. For simplicity of illustration, we have postulated in Fig.~$\ref{75}$ a structure of bigdata aware wireless network, consisting of several mutually complementary components that enable data-driven mobile services, whose functionalities are described below.

\begin{itemize}
  \item \emph{Data-aware cache management: } For quick access under high traffic volumes, cached contents need to be carefully categorized, compactly organized and timely updated. Many types of content objects, such as music and video files, are embedded with \emph{metadata labels} that describe the properties of the contents, from which the data contents could be well classified. By classifying data into a number of sub-classes based on contents, such as sports videos and news pictures, the LUs could achieve more accurate evaluation of the content popularity by jointly considering its own access count and the total access count of its type, which reflects the average frequency of potential future accesses. Accordingly, popular contents are continuously cached while unpopular contents are removed regularly to maximize the effective system bandwidth given limited cache size.
  \item \emph{Crowd computing: } Mobile users of similar interests could share their resources with peers in their vicinity, either with or without taking advantage of the wireless infrastructure. For instance, a complete 3D street view could be generated by a BS from relevant photos contributed by users from different angles. Meanwhile, when MT-to-BS connection is unavailable, an MT could ask for assistance from its neighboring MTs to share available contents and applications, or to even act as relays to the cellular network, etc. Such an idea is explored in \cite{2015:Guo}, where a crowd-enabled data transmission mechanism is proposed to let mobile users assist the data dissemination of other users. In particular, it makes use of personal social information and market incentives to enhance the ``willingness" of mobile users for acting as a data broker of others such that higher chance of successful data delivery could be achieved. Essentially, this peer-to-peer nature of crowd computing exploits user mobility and spatial correlation of data traffics, which also helps us reduce the conventional cellular traffic to and from the wireless infrastructure.
  \item \emph{Mobile cloud processing: } Multiple interconnected C-RANs constitute a mobile cloud, which could optimize the wireless services based on knowledge with respect to mobile traffic patterns, especially when user mobility spans across different C-RAN clusters. For instance, based on the mobility pattern of an MT, a CU could reserve channel resource in advance and pre-feed the contents to the BSs along the anticipated MT's route. As such, chunks of contents could be sent from different BSs to achieve seamless handoffs. Similarly, aggregate characteristic behavior of data traffic could also be used to allocate resources such as bandwidth and cache space to some popular locations ahead of some real-time events. This approach could evidently reduce connection time, delay jitter, and burden of real-time traffic bursts on both cellular fronthaul and backhaul.
  \item \emph{Wireless cloudlet: } The concept of cloudlets introduced in \cite{2009:Satyanarayanan} defines a self-organized light cloud with limited storage and computing power installed at the BSs to enhance their local data processing capability. The deployment of cloudlets could effectively reduce the packet round-time delay by an order of magnitude. A cloudlet may be owned by the network operator but leased to commercial clients for improving performance of delay-sensitive applications, such as online gaming. Besides, a cloudlet could also allow commercial clients to access local cache to provide better location-based services. For instance, an advertising company could send to its subscribers in the vicinity the latest deals based on the information posted by local stores and queries made by prospective customers. With cloudlet, real-time traffic in the backhaul network could also be largely reduced, since many services could be provided locally instead of burdening the core network.
  \item \emph{Context/social-aware processing: } Context/social-aware computing is an emerging paradigm for exploiting complex data characteristics besides conventional user profiles such as mobility pattern and demand distribution \cite{2012:Lukowicz}. The idea of context-aware computing is to provide personalized services adaptive to the MT's real-time ``context", such as traveling, working, and recreation, either directly reported by the MT or inferred from various available data. Social computing, on the other hand, calls for wireless resource allocation to follow closely the interaction within and among social groups \cite{2012:Lukowicz}. Conceptually, a social group is a subset of users that share some similar interests, professions, hobbies, and life experiences, etc. In general, a social group has unique ``eigenbehaviors", such that the group members require and generate similar data contents. The knowledge of a social community's composition, activities and interests could be used to improve the wireless services for the targeted social group members.
  \item \emph{Software-defined-network (SDN):} SDN replaces the conventional hardware-configured routing and forwarding devices by software programmable units. In particular, it decouples the user's data plane (U-plane) from the control and management plane (C-plane), such that the network is managed by a central controller while the underlying devices are only responsible for simple functions such as packet forwarding. Such decoupling provides unprecedented flexibility to network traffic management, where packet forwarding decisions may now be programmed based on many new considerations such as QoS (quality of service) requirement, application types, and payload length, in addition to the conventional destination oriented and distance-based metrics. For SDN-enabled wireless networks, \cite{2014:Liao} proposes a flow-based resource management framework in C-RAN, where the packet routing in the backhaul network and beamforming design in the wireless access network are jointly optimized based on individual data flow's source-destination pair, wireless channel condition, backhaul link capacity, and user QoS requirements, etc. In the case of WLAN networks, \cite{2012:Suresh} introduces a SDN-based enterprise WLANs framework named Odin, which is built with programable functions, global knowledge of network status, and direct control of network devices. The SDN-based system makes many difficult or costly tasks in conventional WLANs easier and less inexpensive, including seamless user handoffs, global load balancing, and hidden terminal problem mitigation.
\end{itemize}

\section{Future Research Directions}\label{sec:future}
In the mobile bigdata era, wireless system designs contain rich research problems of important applications and impact that are yet to be studied. Beyond the many research issues that arise among the number of topics we have discussed so far, here in this section, we highlight several interesting research topics that we particularly find exciting.

\subsection{Reduced-complexity fronthaul processing}
In many data compression proposals, real-time calculation of the optimal compression noise covariance matrix is often impeded by the large number of fronthaul capacity constraints and the non-convex nature of many fronthaul-constrained problems. The problem is further exacerbated by the difficulty in generating practical joint compression codebooks based on the obtained covariance matrix. Therefore, sub-optimal but practical compression schemes, such as scalar quantization, should be given more consideration in future study of fronthaul-constrained compression design. Similarly, CU-level encoding and decoding also suffers from high computational complexity on large-scale multi-user detection and the combinatorial nature of many limited cooperation schemes, such as optimal antenna, relay, modulation and coding combinations, as well as BS selections. It therefore calls for practical complexity-reduction algorithms that are truly scalable to the number of mobile users and network entities.

\subsection{Cache-assisted wireless resource allocation}
BS-level caching is expected to play an important role in future wireless bigdata processing, due to its simplicity, low cost, and natural integration with bigdata analytical tools. However, research on cache-assisted wireless resource allocation is still in its infancy. For cache-assisted cellular networks with BS-level caching, currently there is a shortage of both concrete theoretical analysis on the capacity gain of cache-assisted processing and practical optimization frameworks for cache-assisted resource allocation. Furthermore, effective and optimized integration of various identified bigdata characteristics in cache-assisted network design is an interesting problem that awaits future investigations.

\subsection{Distributed network traffic control}
In large-scale wireless networks, distributed control/computing algorithms could be integrated to alleviate computational complexity of the CU, to reduce backhaul traffic volume and to mitigate the risk of single point failures without compromising overall system performance. Owing to the programmability of SDN-enabled system infrastructure, distributed control mechanisms could be implemented with much better flexibility and lower cost. However, the feasibility and complexity reduction of distributed algorithms are often constrained by the underlying problem structure, such as the coupling constraints in the backhaul and the partial knowledge of data traffic, etc. Distributed control, or a mixed centralized and decentralized control framework, is a promising working direction towards a future wireless networking design supporting mobile bigdata. Additionally, the SDN-based design may also incorporate distributed caching (at BSs and routers) to enhance the efficiency of the routing decision.

\subsection{Mobile data security and privacy}
Harvesting over large mobile data sets and data analytics naturally give rise to concerns with respect to data security and privacy. In a cloud-based wireless network, large amount of data is stored in the fronthaul/backhaul network either for customers' personal use or as commercial database for future analytical purposes. The system operators or commercial entities that collect the user data should be responsible for data security and privacy. For example, personal data should be only available for legitimate and authenticated users. Similarly, data integrity should be guaranteed such that no data is lost or modified by unauthorized entities. Furthermore, it is also important to maintain confidentiality of user data when they are either in storage or during processing. It is therefore important to develop secure yet efficient data processing and storage methods. Promising security measures may include privacy aware distributed data storage and decentralized processing, which aim to maintain local data confidentiality.

\section{Conclusions}\label{sec:conclusion}
This article addresses challenges and opportunities that we face in the era of wireless big data. We first reviewed state-of-the-art signal processing methods and networking structures that may allow us to effectively manage and in fact take advantage of wireless bigdata traffic. We outlined the major obstacles of bigdata signal processing and network design with respect to the scale of problem size and the complex problem structures. Nevertheless, research on big data for wireless communications and networking is not only promising but also inevitable in light of the continuing data volume explosion. We also suggested several interesting research problems aimed at stimulating future wireless research innovations in the bigdata era.

\end{document}